\documentclass{emulateapj}
\newcommand{\ga} {\gtrsim}
\newcommand{\la} {\lesssim}
\lefthead{GAUDI} \righthead{TIDAL CAPTURE OF PLANETS}
\def\au{\rm AU}
\def\rjup{R_{\rm Jup}}
\def\mjup{M_{\rm Jup}}
\def\kms{{\rm km\,s^{-1}}}

\def\au{{\rm AU}}

\def\gyr{{\rm Gyr}}
\def\days{\rm d}

\def\mave{\langle M\rangle}
\def\acap{a_{\rm cap}}
\def\msun{M_\odot}
\def\rsun{R_\odot}
\def\pcpc{\rm pc^{-3}}
\def\rhm{r_{hm}}
\def\drv{{\rm d}}

\newcommand\eq[1]{eq.~(\ref{eqn:#1})}

\begin{document}

\title{Extremely Close-In Giant Planets from Tidal Capture}

\author{B.\ Scott Gaudi\altaffilmark{1}}
\affil{Institute for Advanced Study, Princeton, NJ 08540}
\altaffiltext{1}{Hubble Fellow}
\email{gaudi@ias.edu}

\begin{abstract}

Planets that form around stars born in dense stellar environments such
as associations, open clusters, and globular clusters are subject to
dynamical perturbations from other stars in the system.  These
perturbations will strip outer planets, forming a population of
free-floating planets, some of which will be tidally captured before
they evaporate from the system.  For systems with velocity dispersion
$\sigma \sim 1~\kms$, Jupiter-mass planets can be captured into orbits
with periods of $0.1-0.4$ days, which are generally stable over $\sim
10^9$ years, assuming quadratic suppression of eddy viscosity in the
convective zones of the host stars.  Under this assumption, and the
assumption that most stars form several massive planets at separations
$5-50~{\rm AU}$, I estimate that $\sim 0.03\%$ of stars in rich,
mature open clusters should have extremely close-in tidally captured
planets.  Approximately $0.005\%$ of field stars should also have such
planets, which may be found in field searches for transiting planets.
Detection of a population of tidally-captured planets would indicate
that most stars formed in stellar clusters.  In typical globular
clusters, the fraction of stars with tidally-captured planets under
these assumptions rises to $\sim 0.1\%$ -- in conflict with the null
result of the Hubble Space Telescope transit search in 47 Tuc.  This
implies that, if the quadratic prescription for viscosity suppression
is correct, planetary formation was inhibited in 47 Tuc: on average
$\la 1$ planet of mass $\ga \mjup$ (bound or free-floating) formed per
cluster star.  Less than half of the stars formed solar-system
analogs.  Brown dwarfs can also be captured in tight orbits; the lack
of such companions in 47 Tuc in turn implies an upper limit on the
initial frequency of brown dwarfs in this cluster.  However, this
upper limit is extremely sensitive to the highly uncertain timescale
for orbital decay, and varies by four orders of magnitude depending on
the choice of prescription for the suppression eddy viscosity.
Therefore, it is difficult to draw robust conclusions about the
low-mass end of the mass function in 47 Tuc.

\end{abstract}

\section{Introduction\label{sec:intro}}

The discovery of an extrasolar planetary companion orbiting a mere
$0.05~\au$ from its host star 51 Peg \citep{mandq95} originally came as
a surprise, given the seeming impossibility of forming a giant planet
so near to its parent star.  However, the theorists quickly recovered,
and, drawing on older works which studied interactions between protoplanets
and their natal disks \citep{gandt80,ward86}, invoked planetary
migration as the mechanism for delivering the companion to 51 Peg from
its birthplace to its current observed position (e.g.,
\citealt{lbr96}).  

Since this time, considerable progress in the
theory of planetary migration, combined with the detection of a large number of
additional giant planets on short-period orbits, has left little doubt
that this process is important in shaping the architecture of
planetary systems.  However, many questions remain.  For example, the
mechanisms by which planetary migration is stopped are uncertain
\citep{lbr96,trilling98,kandl02}.  These mechanisms have the additional
burden that they should explain both the observed `pile-up' of planetary companions at
periods of $P=3~\days$ (i.e.\ \citealt{kandl02}), and the newest
discovery of a planet at $P=1.2~\days$ \citep{konacki03,sasselov03}.
In addition, it is clear from examples of extrasolar giant planets
orbiting close to their supposed birthplaces, as well as our own solar
system, that orbital migration is not always so efficient.  A
satisfactory explanation for the causes of these variations in
the efficiency of migration is lacking. 

It seems plausible that many, if not the majority, of disk stars were
initially formed in stellar systems -- loose associations or open
clusters that have since dissolved.  Evidence for this comes in part
from studies of the nearby moving groups \citep{dezeeuw99}.  Planets
orbiting stars in stellar systems are subject to numerous effects
resulting from the dense stellar environment.  Planets orbiting
at larger distances from their parent stars are generally more sensitive
to these effects.  Therefore, the structure of planetary
systems will be the result of a complex interplay between the local
properties of the system that drive planetary formation and migration, and 
the non-local effects that arise from the star's environment. 
Thus, if it is indeed the case 
that most stars formed in stellar clusters, planetary systems cannot be
understood as isolated systems: a complete picture of planetary formation and
evolution must consider both local and non-local effects.

Several authors have considered the effects of dense stellar environments
on the formation and survival of planetary systems
\citep{armitage00,ds01,sb01,bsdh01,hs02}. From these studies it is
clear that, if distant planets can form in stellar clusters, they are
likely to be stripped from their parent stars, resulting in a
free-floating population of planets that will slowly evaporate from
the system.  Recently, \citet{bonn03} suggested that freely-floating
brown dwarfs (BD) in globular clusters will occasionally be tidally
captured, resulting in a detectable population of extremely close-in BD
companions to main-sequence stars if the frequency of freely-floating
BDs is large. Similar considerations can also be applied to planets 
in stellar systems.

Here I consider the formation and stability of extremely close-in
giant planetary companions to stars via tidal capture of free-floating
planets.  In \S\ref{sec:dynamical}, I summarize the dynamical
processes of ionization, evaporation, capture, and tidal decay, and
their relevant timescales.  In \S\ref{sec:frequency}, I give a crude
estimate for the frequency of extremely close-in giant planets as a
function of the parameters of the stellar system.  In \S\ref{sec:bds},
I consider tidally-captured BD and planetary companions to stars in
globular clusters, and interpret the null result of the 47 Tuc transit
search \citep{gilliland00}. Finally, in \S\ref{sec:implications}, I
discuss possible implications and prospects for the detection of
such a population of planetary companions.

\section{Dynamical Processes\label{sec:dynamical}}

It is not clear if planets can form around stars in stellar systems at
all.  In particular, in rich clusters with $N\ga 10^3$ members,
high-mass stars can generate an ultraviolet radiation field that is
sufficiently intense to photoevaporate protoplanetary discs in a few
hundred thousand years \citep{armitage00}.  However, planet formation
might still be possible if there exists a substantial delay between
high and low-mass star formation.  This fact, combined with our
incomplete understanding of the efficiency of planetary migration,
make it difficult to predict a priori the frequency and distribution
of planets orbiting stars in stellar systems.  For the purposes of 
discussion, I will sweep all of these uncertainties under the rug, and
simply consider the evolution and consequences of a substantial
population of long-period planets around stars in stellar systems.

Binaries in stellar systems evolve under random encounters with other stars
in the system.  This evolution proceeds according to Heggie's law \citep{heggie75},
such that soft binaries tend to get disrupted by encounters.  A binary is
soft when its binding energy is equal to the mean kinetic energy of the
stars in the system.  Thus there is a critical separation above which
binaries will tend to be disrupted,
\begin{equation}
a_{\rm dis}= \frac{G m_p M_*}{2\mave\sigma^2},
\label{eqn:adis}
\end{equation}
where $m_p$ is the mass of the secondary, $M_*$ is the mass of the primary,
$\mave$ is the average mass and $\sigma$ is the velocity dispersion of the
cluster stars, respectively.  Planetary companions to solar-type
stars ($m_p\sim \mjup$ and $M_*\sim \msun$) in typical open clusters
($\mave \sim 0.5 \msun$ and $\sigma \sim 1.5\kms$ are will tend to 
get disrupted if $a\ga 0.4\au$, or $P\ga 100\days$.  

The timescale for disruption of a planet is approximately the time it takes
for the random encounters to change the planet's energy by an amount equal
to its binding energy \citep{bandt87},
\begin{equation}
t_{\rm dis} = (1+q)\frac{M_*}{\mave}\frac{\sigma}{16\sqrt{\pi}G\mave \nu a \ln \Lambda}.
\label{eqn:tdis}
\end{equation}
Here $q\equiv m_p/M_*$, $\nu$ is the stellar number density, and $\ln
\Lambda \simeq \ln 0.4 N$ is the Coulomb logarithm.  The number
density and velocity dispersion within a stellar system can vary by several
orders of magnitude.  For simplicity, here I will 
adopt these properties averaged over the region interior to $\rhm$, the
half-mass radius of the cluster.  The mean stellar density in this
area is simply
\begin{equation}
\nu=\frac{\frac{1}{2}N\mave}{\frac{4}{3}\pi\rhm^3}.
\label{eqn:numean}
\end{equation}
For a wide range of
equilibrium stellar systems, the half-mass radius can be related to
the total mass and velocity dispersion of the system by \citep{spitzer87},
\begin{equation}
\rhm = \frac{0.4G\mave N}{3\sigma}.
\label{eqn:rhm}
\end{equation}
I will adopt this relation throughout unless stated otherwise.  For an
average stellar density $\nu \sim 10^3~\pcpc$ and velocity dispersion
$\sigma \sim 1.5~\kms$, the above relation yields $N\sim 7600$.  For
the stellar parameters given above, $t_{\rm dis} \sim 0.5~\gyr
(a/2.5\au)^{-1}$.  Thus most planets with $a \ga 2.5 \au$ in an open
cluster will be disrupted over the cluster's lifetime ($\sim
0.5~\gyr$).

Since planets are typically liberated due to the accumulated effects
of distant encounters, free-floating planets will initially have
velocities similar to the stars, and will not escape the system
immediately \citep{hs02}.  However, equipartition of energy will slowly
drive the velocity dispersion of free-floating planets to the point
where many of the planets have sufficient velocities to
evaporate from the system entirely.  Assuming that the stellar
component of the system dominates the mass, the characteristic timescale for
this process is the relaxation timescale \citep{spitzer87,bandt87},
\begin{equation}
t_{\rm rlx} = 0.34 \frac{\sigma^3}{\nu (G\mave)^2 \ln \Lambda}.
\label{eqn:relax}
\end{equation}
For the parameters adopted above, this yields $t_{\rm rlx} \sim 0.03~\gyr$.

\begin{figure}[htbp]
\plotone{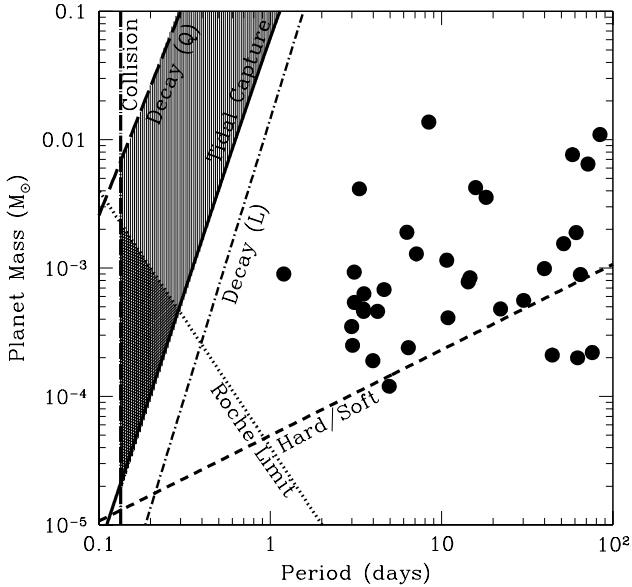}
\caption{ The lines indicate regions in the plane of planet mass
and period where various processes are important, assuming solar-mass,
solar-radius primaries.  Planets can be captured into orbits the left
of the solid line.  Planets captured to the left of the
long-dashed-dot line result in mergers.  Planets to the left of the
dashed (short-dashed-dot) line are stable over $0.5~\gyr$, assuming
quadratic (linear) suppression of eddy viscosity.  Planets to the left
of the dotted line have radii that exceed their Roche-lobe, assuming a
planetary radius of $\rjup$.  Planets to the right of the dashed line
residing in a cluster with velocity dispersion of $\sim 1.5\kms$ are
soft, and tend to be disrupted by perturbations from other cluster
stars.  The lightly-shaded region indicates planet parameters than can
result in stable captures, assuming quadratic viscosity suppression.
The solid points show the mass and period known extrasolar planetary
companions.  }
\label{fig:one} 
\end{figure}

Freely-floating planets will occasionally pass sufficiently close to
another star to raise a significant tide.  If the energy required to
raise this tide is larger than the relative energy of the planet and
star at infinity, then the encounter will lead to a bound system
\citep{fabian75,press77,lee86}.  This leads to the requirement that the
planet must pass within a minimum distance $\acap$ to be captured
\citep{fabian75},
\begin{equation}
\acap \simeq R_* \left[ \frac{GM_*}{R_* \sigma^2} q(1+q)\right]^{1/6}.
\label{eqn:acap}
\end{equation}
For $q=10^{-3}$, $M_*\sim \msun$, $R_*\sim \rsun$, $\acap/R_* \sim 2$.  Thus the planet
must pass within two stellar radii to be captured.

The timescale for tidal capture is,
\begin{equation}
t_{\rm enc}\simeq \frac{\sigma}{8 \sqrt{\pi} GM_* \nu \acap},
\label{eqn:tenc}
\end{equation}
where I have assumed that $GM_*/2\sigma^2 \simeq a_{\rm dis}/q \gg
\acap$, which is valid for the cases considered here.  I find 
$t_{\rm enc} \simeq 500~\gyr (\acap/2\rsun)^{-1}$.  Assuming every star forms
$N_{p}$ planets that are liberated from their parent stars essentially
immediately, the fraction of stars with
tidally captured planet can be crudely estimated as $\sim N_{p}
(t_{\rm rlx}/t_{\rm enc}) \sim 0.01\% N_{p}$. 

Planets passing within $a_{\rm coll} = R_p+R_*$ of a star will
physically collide with it, resulting in a merger.  This
therefore sets an absolute lower limit on the period of a tidally
captured planet.  For gaseous objects supported only by degeneracy
pressure (including giant planets and brown dwarfs), $R_p\sim
0.1\rsun$, roughly independent of mass.  For $R_*\sim \rsun$, $a_{\rm
coll} \simeq 0.005\au$, or a period of $P\simeq 0.13~\days$.  This also
sets a lower limit on the mass of a planet that can be captured, as
very low-mas planets must pass within the collision separation in
order to raise a sufficient tide to be captured.  This limit is found
by equating $a_{\rm coll}$ with $\acap$:
\begin{equation}
q \ga \frac{R_*\sigma^2}{GM_*}\left(1+\frac{R_p}{R_*}\right)^6,
\label{eqn:qmin}
\end{equation}
where I have assumed $q\ll 1$.  For the fiducial parameters, this is $q \ga 2\times 10^{-5}$. 

Low-mass planets may also be tidally captured sufficiently close to
their parent stars that their radii exceed the Roche limit.  The
limiting separation for Roche lobe overflow is,
\begin{equation}
a_{\rm R}=\left(\frac{3}{q}\right)^{1/3}R_p\simeq 1.44\left(\frac{q}{10^{-3}}\right)^{-1/3}\rsun,
\label{eqn:roche}
\end{equation}
where, for the last relation, I have assumed $R_p=0.1\rsun$.  The
Roche limit separation exceeds $a_{\rm coll}$ for $q\ga 2 \times
10^{-3}$.  Planets that are captured within their Roche limit will
lose mass to their parent star.  This mass loss is accompanied by a
transfer of angular momentum, resulting in an outward migration of the
planet.  Thus a planet initially captured inside its Roche limit
can still result in a stable system.  The lower the mass of
the planet, the farther it must migrate to halt mass transfer.  Very
low mass captured planets may not survive at all.

Planets captured into very close orbits will quickly circularize, on a
timescale of $\sim 10^5$ years \citep{rasio96}, and thereafter be
subject to tidal decay.  The timescale for tidal decay is (e.g.\
\citealt{rasio96}),
\begin{equation}
t_{\rm dec} = \frac{f}{t_{\rm c}}\frac{M_{\rm cz}}{M_*} q(1+q)\left(\frac{R_*}{a}\right)^8,
\label{eqn:tdec}
\end{equation}
where $M_{\rm cz}$ is the mass of the convective zone; for solar-type
stars, $M_{\rm cz}\sim 0.02M_\odot$.  Here $t_{\rm c}$ is the
convective eddy turnover timescale, which I will approximate as
$t_{\rm c} \sim [M_{\rm cz} 0.2R_*^2/3L_*]^{1/3}$, where $L_*$ is the
luminosity of the star (see, e.g., \citealt{rasio96}).  For solar-type stars, $t_{\rm
c} \sim 20~\days$.  Thus for tidally-captured planets, $P\ll t_{\rm
c}$, and the largest eddies can no longer contribute to the total
viscosity, resulting is a suppression of the viscous dissipation.
Therefore $f$ is less than unity.  However, the correct form for $f$
remains controversial (see \citet{goodman97} for a discussion).
Following \citet{sasselov03}, I will consider both linear suppression
of the eddy viscosity \citep{zahn89},
\begin{equation}
f_{L}=\left( \frac{P}{2t_{\rm c}}\right),\qquad P\ll t_{\rm c},
\label{eqn:flin}
\end{equation}
as well as quadratic suppression \citep{gk77},
\begin{equation}
f_{Q}=\left( \frac{P}{2\pi t_{\rm c}}\right)^2, \qquad P\ll t_{\rm c}.
\label{eqn:fquad}
\end{equation}
Although quadratic suppression is theoretically better motivated,
linear suppression appears to be in better agreement with the observed
timescale of tidal circularization in close binaries \citep{goodman97}.

For tidally-captured planets with $\acap \simeq 2\rsun$, the decay
timescale for linear suppression is $t_{\rm dec} \simeq 0.1~\gyr$,
whereas for quadratic suppression, $t_{\rm dec} \simeq 80~\gyr$.
Therefore the choice of prescription has an enormous effect on the
number of surviving tidally-captured planets: one expects a negligible
number of surviving planets for linear suppression,
whereas planets are stable over much longer than $\sim \gyr$
for quadratic suppression.

The angular momentum lost from the tidal decay of the planet's orbit will be transfered
to the star, spinning it up.  Assuming the star is initially slowly
rotating, spin-up of the star will occur on a timescale \citep{rasio96},
\begin{equation}
t_{\rm su}\sim \frac{I_*}{I_{orb}}t_{\rm dec}=
\frac{k^2}{q}\left(\frac{R_*}{a}\right)^2 t_{\rm dec},
\label{eqn:tsu}
\end{equation}
where $I_*$ and $I_{orb}$ are the moments of inertia of the star and orbit,
and $k^2\equiv I_*/M_*R_*^2$.   For $a=a_{\rm cap}\sim 2R_*$, and $k^2=0.08$,
$t_{\rm su}\sim 20 (10^{-3}/q) t_{\rm dec}$.  Thus companions with $q\la 0.02$ will
decay before they spin-up the star sufficiently to synchronize the star's
spin period with the orbital period.
Conversely, companions with $q\ga 0.02$ will synchronize before they decay completely.
This equilibrium is stable only if the orbital angular momentum of the planet
is more than three times the spin angular momentum of the star \citep{hut80}.  This
leads to a minimum mass ratio for stability,
\begin{equation}
q\ge 3 k^2 \left(\frac{R_*}{a}\right)^2.
\label{eqn:qstable}
\end{equation}
Thus companions with $q\la 0.06$ cannot reach a stable equilibrium,
and will always decay.  More massive companions may reach a stable
equilibrium.  

The question of orbital stability via synchronization is further
complicated by the fact that the star will likely be spun-up as a
result of the capture process itself\footnote{I am indebted to I.\
Bonnell for pointing this out.}, and thus may be fairly rapidly
rotating.  This will not only affect the viscous dissipation
mechanisms, but will also generally shorten the timescale for
synchronization.  Given the uncertainties involved in determining the
effects of these various process, and since I am primarily interested
in planetary-mass companions, for which synchronization is likely
less important, I will furthermore ignore synchronization entirely, and assume
that all orbits decay on the timescale given by \eq{tdec}.
However, I note that this issue has direct bearing on the discussion
of the stability of tidally-captured BD companions to stars in 47 Tuc
in \S\ref{sec:bds}.

Figure \ref{fig:one} summarizes the regions of parameter space in the companion
mass-period plane where ionization, tidal capture, and tidal decay 
are important, as discussed in this section.  
For this figure, I have
assumed as before $M_*=\msun$, $\mave=0.5\msun$, $R_*=\rsun$,
$\sigma=1.5~\kms$, and $\nu=10^3~\pcpc$.  In addition, I have assumed a
fixed companion radius of $R_p=\rjup$, and a system lifetime of
$0.5~\gyr$.  Assuming quadratic suppression prescription of tidal decay, there is a
small sliver of parameter space in the period-mass plane of where
planets can be tidally captured and are stable to tidal decay over
typical open cluster lifetimes.  For linear suppression, there is no region
of parameter space where tidally-captured planets are expected to be stable 
on such timescales.
Also shown are known extrasolar
planets.  Clearly tidally-captured planets occupy a region of
parameter space that is relatively disjoint from known companions, and
would likely to easily identified as the result of capture (rather
than migration).

\section{Frequency Estimates\label{sec:frequency}}

The frequency of tidally-captured planets in a stellar system of a given
age depends on the competing effects of ionization, evaporation, capture,
and orbital decay.  The exact number will depend on the balance
between the timescales of these various dynamical effects.  This will
in turn depend on the dynamical properties of the cluster at all times
in its evolution, and thus the processes of relaxation, mass
segregation, mass loss, etc., should all be considered.  Ideally, the
most robust way to accomplish this is through detailed N-body simulations.
However, such a study is outside of the scope of this paper.
Furthermore, such a detailed study is perhaps not warranted, as the
number of tidally captured planets will also depend critically on the
initial frequency and distribution of planets formed around stars in
the cluster.  Given that a general theory of planet formation is
lacking, this can only be a wild guess at best.  I therefore simply
provide only a crude estimate for the frequency of tidally-captured
planets.  This estimate should be good to an order-of-magnitude, and
should serve to elucidate the dependence of the frequency on the
gross parameters of the stellar system and the input assumptions.

I consider only average properties of the stars and planets in
the stellar system.  The dynamical properties of the stellar system
are specified by the average number density $\nu$ and velocity
dispersion $\sigma$ inside the half-mass radius, and the average mass
of the stars, $\mave$.  The total number is then set by equations
(\ref{eqn:numean}) and (\ref{eqn:rhm}).  Unless otherwise stated, I
assume $\mave=0.5 \msun$,  which is roughly appropriate for mass
functions observed in the field, young clusters, open clusters,
and globular clusters (see \citealt{chabrier03}).  
Here and throughout, I adopt radii and bolometric luminosities from
\citet{allen76} and convective zone masses from \citet{pdc01}.

For my fiducial calculations, I assume that every star in the system
originally has four companions with mass $M_p=\mjup$, with a
distribution that is uniform in $\log{a}$ between $5-50\au$.  This
choice is primarily motivated by the distribution of massive planets
in our own solar system, but is not inconsistent with the distribution
of extrasolar planets detected via radial velocity surveys.  These
surveys are only just becoming sensitive to planets around solar-type
stars at $5\au$; however extrapolations based on current samples
indicate that the fraction of stars with Jupiter-mass planets at
periods longer than this may be quite large (e.g., \citealt{lg03}).  For
the same distribution of $a$ but other choices for the frequency
of massive planets, one can simply scale all results by $(N_p/4)$,
where $N_p$ is the average number of planets per star.

\begin{figure}[htbp]
\plotone{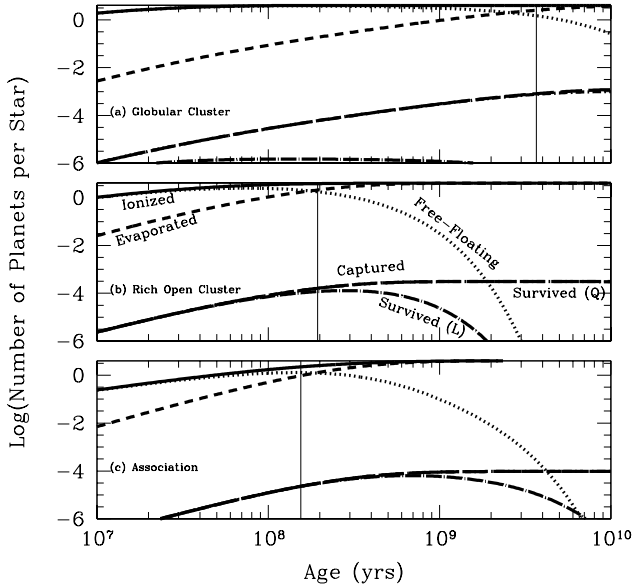}
\caption{ Evolution of the planetary population of stellar clusters.
Every star in the cluster is assumed to initially have $N_p=4$ planets
with a uniform logarithmic distribution between 5 and 50 AU.  For
other values for the number of planets per star, all curves can be
scaled as $N_p/4$.  Each curve shows the number of planets per star as
a function of the age of the cluster.  The solid curve shows the
number of planets ionized from their parent stars.  The dashed curve
shows the number of these planets that have escaped the cluster;
dotted curve shows the remaining free-floating population.  The
long-dashed curve shows the number of tidally captured planets.  The
short-dashed-dot (long-dashed-dot) curve shows the fraction of tidally
captured planets that have survived orbital decay assuming quadratic
(linear) suppression of eddy viscosity. The vertical line shows
$t_{\rm rlx}/\xi_e$, the typical timescale for evaporation from the
system. (a) Assuming a cluster velocity dispersion of $\sigma=10~\kms$
and a number density of $\nu=10^4~\pcpc$, appropriate to a globular
cluster.  (b) $\sigma=1.5~\kms$ and $\nu=10^3~\pcpc$, appropriate to a
rich open cluster.  (c) $\sigma=0.6~\kms$ and $\nu=10^2~\pcpc$,
appropriate to a loose association.  }
\label{fig:two} 
\end{figure}

Free-floating planets in stellar stellar are depleted via evaporation and
replenished by disruption of bound planets\footnote{Free-floating planets
are also depleted via captures, but the rate of captures is generally so small that this
can safely be ignored.}.    The rate of change in the number $N_{\rm ff}$
of free-floating planets per star is thus
\begin{equation}
\frac{\drv N_{\rm ff}}{\drv t}= \frac{\drv N_{\rm dis}}{\drv t} - 
\frac{\drv N_{\rm evap}}{\drv t}.
\label{eqn:dnffdt}
\end{equation}
Here $\drv N_{\rm dis}/\drv t$ is the rate at which planets are ionized,
\begin{equation}
\frac{\drv N_{\rm dis}}{\drv t}=\frac{\drv }{\drv t}\left[\int_0^{\infty} \frac{\drv N_p}{\drv \log a} {\cal P}_{\rm dis}(a) \drv \log a\right].
\label{eqn:dnioindt}
\end{equation}
Here ${\drv N_p}/{\drv \log a}$ is the initial distribution of
planetary companions, which is constant between $5-50~\au$ and zero
otherwise, and ${\cal P}_{\rm dis} \equiv 1- e^{-t/t_{\rm dis}(a)}$ is
the probability that planet with separation $a$ will be disrupted.
The rate at which planets are evaporated is
\begin{equation}
\frac{\drv N_{\rm evap}}{\drv t}=\frac{\xi_e N_{\rm ff}}{t_{\rm rlx}},
\label{eqn:dnevapdt}
\end{equation}  
where $\xi_e=0.156$ is the evaporation probability appropriate for
test particles \citep{spitzer87}.  The number of free-floating planets
as a function of time can be found by integrating \eq{dnffdt} from
$t=0$.

Tidally-captured planets are depleted via orbital decay and
replenished via new captures of free-floating planets.  The rate of
change in the number $N_{\rm sur}$ of tidally-captured planets per
star is thus
\begin{equation}
\frac{\drv N_{\rm sur}}{\drv t}= \frac{\drv N_{\rm cap}}{\drv t} - 
\frac{\drv N_{\rm dec}}{\drv t}=\frac{N_{\rm ff}}{t_{\rm enc}}-\frac{N_{\rm sur}}{t_{\rm dec}}.
\label{eqn:dnsurdt}
\end{equation}
I determine the mean capture cross section by averaging the width of
the region between $\acap$ and $a_{\rm coll}$, which depends on the
relative velocity between the planet and star, over a Maxwellian
(relative) velocity distribution with dispersion $\sigma$.  I then
evaluate $t_{\rm enc}$ for this cross section, and $t_{\rm dec}$ for
the average semi-major axis of the captured planets.  The number of
surviving free-floating planets is then found by integrating
\eq{dnsurdt}.

Note that I have ignored the effects of mass segregation.  Mass
segregation will decrease the rate at which planets are
captured, as it will partially decouple the dynamical
interactions between the stellar and free-floating planetary
populations.  This will generally be a relatively small correction, as
most planets are tidally captured during the first few
relaxation times of the cluster. 

Figure \ref{fig:two} shows the resulting evolution of the planetary
population for three types of stellar systems, characterized by the
parameters $\nu$ and $\sigma$.  These are $\nu=10^4~\pcpc$ and
$\sigma=10~\kms$, appropriate to a globular cluster, $\nu=10^3~\pcpc$,
$\sigma=1.5~\kms$, appropriate to a rich open cluster, and
$\nu=10^2~\pcpc$, $\sigma=0.6~\kms$, appropriate to a loose association.
These systems have a total number of stars and half-mass radii equal
to $N\simeq 7\times 10^5$, $\rhm=2~{\rm pc}$ (globular cluster),
$N\simeq 7600$, $\rhm\simeq 1~{\rm pc}$ (rich open cluster) and $N
\simeq 1500$, $\rhm \simeq 1.2~{\rm pc}$ (association).  These values
are not meant to be definitive, but merely span the interesting range
of possible systems. Note that $\rhm$ does not enter into the
calculations directly, and $N$ enters only logarithmically via the
Coulomb logarithm.  

Due to their high stellar density and velocity dispersions, globular
clusters generally have relatively small disruption timescales and
long relaxation timescales.  Therefore, planets are liberated from
their parent stars very quickly, and evaporation is slow, making the
dynamics particularly simple: the number density of free-floating planets 
is roughly constant, and $\drv \log{N_{\rm sur}}/\drv \log t$ is approximately
constant when tidal decay is negligible.  At the typical globular cluster lifetime
of $10~\gyr$, $N_{\rm sur}\simeq 0.1\%$.  

The evolution of rich open clusters is more complicated due to their
short relaxation times.  From inspection of Figure \ref{fig:two}, it
is clear that the majority of planets are tidally captured within a
time $t_{\rm rlx}/\xi_e \sim 0.2~\gyr$, after which the number of surviving
planets is approximately constant at $N_{\rm sur}\simeq 0.03\%$.  The lifetime of open clusters
is uncertain, but rich clusters have been observed with ages of a $\gyr$ or more 
\citep{kalirai01,burke03a}.  Less massive clusters are probably
dissolved on shorter timescales. 

Loose associations have even shorter relaxation times, but are
otherwise similar to open clusters.  However one important difference
is that, because of their smaller velocity dispersions, planets are
captured at larger separations.  Due to the extremely strong
dependence of the tidal decay rate on the semi-major axis, this has a
profound effect on the survival of tidally-captured planets for the
linear prescription for the suppression of eddy viscosity: even under
such a prescription, one expects a significant number of surviving
tidally-captured planets at a $\gyr$ or less.  Loose associations are
likely to disperse fairly rapidly, on timescales of a few hundred
million years or less.  At $10^8$ years, $N_{\rm sur}\simeq 0.005\%$

\section{Brown Dwarfs and Planets in 47 Tuc\label{sec:bds}}

Because of their high encounter rates and large numbers of stars, globular
clusters are expected to contain a large number of systems formed via
tidal capture. Using the results from the previous section, a system
with $\nu \sim 10^4~\pcpc$ and $\sigma \sim 10~\kms$ should have $\sim N
N_{\rm sur}=7\times 10^5 \times 0.001 \sim 700$ systems of
tidally-captured planets.  This process is not limited to planets
stripped from their parent stars, of course: any planets and brown
dwarfs formed in isolation will also be tidally captured by stars in
globular clusters.  The initial mass function of single objects 
is observed to be approximately flat down to well below
the hydrogen burning limit in many young clusters 
(\citealt{chabrier03}, but see \citealt{briceno02}).
Such a mass function implies a similar number density of stars
and brown dwarfs.  If this mass function is universal then globular clusters 
should also contain a large number of stars with tidally-captured BD companions 
\citep{bonn03}.

Because of their extremely close orbits, tidally-captured planetary and BD
companions have a very high probability ($\ga 25\%$) of transiting
their parent star, making photometric monitoring an ideal way of
detecting such companions.  \citet{gilliland00} monitored the globular
cluster 47 Tuc continuously for $\sim 8~\days$ using the Hubble Space
Telescope, but found no transits.  This null result can be used to
place constraints on the initial frequency of planetary companions as
well as freely-floating planets and BDs formed in isolation \citep{bonn03}.

\begin{figure}[htbp]
\plotone{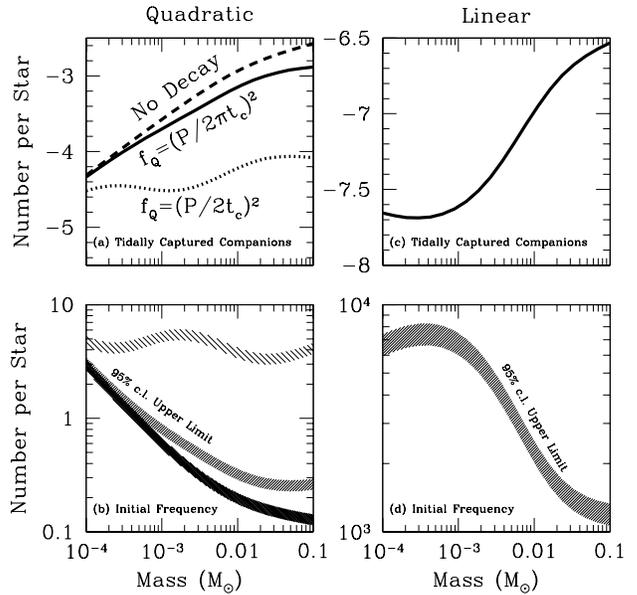}
\caption{ 
(a) Number of tidally-captured companions per star in 47 Tuc
as a function of the mass of the companion in solar masses, assuming
an initial frequency of one per star.  A stellar density of
$\nu=10^4~\pcpc$, velocity dispersion of $\sigma=12~\kms$, half-mass
radius of $\rhm=3.9{\rm pc}$, and an age of $13~\gyr$ were assumed.
The solid curve is for the nominal quadratic assumption of $f_Q=(P/2\pi t_c)^2$ 
for the suppression $f_Q$ of eddy viscosity, whereas the dotted line is
for the assumption $f_Q'=(P/2 t_c)^2$ adopted by some authors.  Here $P$ is the
period of the planet and $t_c$ is the convective eddy turnover time.
The dashed line is the result for no tidal decay. 
(b) 95\% confidence level upper limits to the initial number of
objects (brown dwarfs or planets) per star implied by the null result
of the transit search in 47 Tuc \citep{gilliland00}, as a function of
the mass of the object.  Upper and lower curves bracket the range of
uncertainty in the limits.  See text.  The lower, middle, and upper
curves correspond to the limits from the upper, middle and lower curves
in panel (a), respectively. (c) Same as panel (a), but assuming
linear suppression of eddy viscosity $f_L=(P/2t_c)$.  (d) Same as panel
(b), but for $f_L$. 
}
\label{fig:three} 
\end{figure}

Figure \ref{fig:three} shows the expected number of tidally-captured
companions per star in 47 Tuc as a function of the mass of the
companion, for both linear and quadratic prescriptions for
viscosity suppression.  I have adopted parameters appropriate to 47
Tuc, namely $\nu=10^4~\pcpc$, $\sigma=12~\kms$, $\rhm=3.9{\rm pc}$,
$\mave=0.56\msun$ and an age of $13~\gyr$ \citep{djorgovski93,
gebhardt95, paresce00}.  Note that here I do not assume \eq{rhm}.
Figure \ref{fig:three} assumes an initial frequency of one object
(planet or brown dwarf) per star. Due to the extremely high
disruption rate, these results are essentially independent of whether
the objects are initially free-floating, or bound on orbits with
$a\ga 1~\au$.  For the quadratic prescription for viscosity suppression,
the expected frequency of tidally-captured BD ($0.05\msun$) companions
is $N_{\rm sur}\sim 0.002$ companions per star, whereas the frequency
of tidally-captured Jupiter-mass planets is $N_{\rm sur}\sim 0.0002$.

\citet{gilliland00} claim that, if $f=0.8-1.0\%$ of the 34,091
main-sequence stars they monitored had $1.3\rjup$ companions at
periods of $3.5\days$, then they should have seen 17 planets.  This
implies a 95\% confidence level (c.l.) upper limit of $3f/17=0.14-0.18\%$ to
the fraction of main-sequence stars with such companions.  This null
result can also combined with the prediction for the number of
tidally-captured companions $N_{\rm sur}$ to determine an upper limit
to the initial frequency $N_p$ of objects per star of
\begin{equation}
N_p \le \frac{3}{17} \frac{f}{N_{\rm sur}}\frac{{\cal P}_{T}(3.5\days)}{{\cal P}_{T}(P_{\rm cap})}.
\label{eqn:npmax}
\end{equation} 
Here ${\cal P}_T=(R_p+R_*)/a$ is the transit probability.  Figure
\ref{fig:three} shows the resulting 95\% c.l.\ upper limit to the
initial frequency of objects as a function of the mass of the object.
The shaded region comes from varying $f$ in the range $0.8\%\le f \le
1.0\%$.

The null result implies that, on average, each star in 47 Tuc originally formed
$\le 1$ planet with mass $M_p\ge\mjup$, or $\le 0.4$ planets with
$M_p\ge 5\mjup$.  Less than 50\% of stars in 47 Tuc could have 
formed solar-system analogs.  While this limit is considerably weaker than the
original limit placed by \citet{gilliland00} of $\sim 0.1\%$, it is
important to emphasize that this original limit applies only to planets
in close ($\sim 3~\days$) orbits.  Since planets in such small orbits
were not formed in situ, and must be the result of migration, drawing
conclusions about the formation of planets at larger separations based
on the lack of planets at small separations is subject to
uncertainties in the timescale for orbital migration.  If the
migration timescale were greater than or of order the disruption timescale, the
planets would have been stripped from their parent stars before they had a
chance to migrate sufficiently close to be detectable via transits.  In
contrast, the limit on the frequency of planets implied by the lack of
tidally-captured planets, although weaker, is essentially independent
of where they were initially formed.  In fact, one can conclude the initial
number of free-floating planets (planets formed in isolation via gas fragmentation
and collapse) is also less than or equal to the number of stars.

Under a similar set of assumptions as adopted here, but using a more
sophisticated simulation and neglecting orbital decay, \citet{bonn03}
found an upper limit to the BD frequency in 47 Tuc of $\sim 15\%$.
Figure \ref{fig:three} shows the upper limit derived using the
formalism in \S\ref{sec:frequency}, but neglecting orbital decay.  I find
$12\%$, in good agreement with \citet{bonn03}.  This suggests that, in
the absence of orbital decay, the predicted frequency of
tidally-captured BD companions should be reasonably robust.

For quadratic suppression of eddy viscosity, the timescale for orbital
decay of tidally-captured BD ($q\sim 0.1$) companions is $t_{\rm
dec}\sim 9~\gyr$.  Therefore, approximately $\exp(-9/13)\sim 50\%$ of
the captured BD will still be present today.  Thus tidal decay is
significant at the current age of 47 Tuc for tidally-captured BD
companions.  Figure \ref{fig:three} shows the upper limit including
orbital decay.  The limit is revised upward by $\sim 2$ relative to
the estimate of \citet{bonn03}.  The lack of transiting BD companions
in 47 Tuc implies that the initial frequency of BDs relative to stars
is $\la 25\%$.

The upper limit of $\sim 25\%$ on the relative frequency of BDs in 47 Tuc
is inconsistent with the observed mass functions of most
young clusters and the field,
for which BDs and stars exist in equal numbers 
\citep{chabrier03}, although it is
consistent with the frequency of BDs in the Taurus star-forming 
region \citep{briceno02}.  
At face value, this is an indication that the
initial mass function compact objects is not universal \citep{bonn03}.
Unfortunately, this conclusion relies heavily on the assumed value for
the tidal decay timescale.  Different assumptions for the mechanisms
for tidal dissipation will give rise to radically different
conclusions.  For example, in Figure \ref{fig:three} I show the
predicted frequency of tidally-captured companions under the
assumption of linear suppression of eddy viscosity.  The frequencies
are always $\le 3\times 10^{-6}$, i.e.  $\la 5$ in the entire cluster.
This low number results from the fact that the tidal decay timescale
under linear suppression is smaller by a factor of $\sqrt{f_Q}\sim
10^{-4}$.  Thus $t_{\rm dec}\simeq 4\times 10^6~{\rm yrs}$, and the
frequency of tidally-captured objects is only $\sim t_{\rm dec}/t_{\rm
enc} \sim 10^{-6}$.  Obviously the resulting upper limits are not very
interesting.  Even if one accepts that the quadratic suppression of eddy
viscosity is correct, there are still some ambiguities.  For example,
there exist discrepancies in the quoted form for $f_Q$:
\citet{goodman97} and \citet{sasselov03} use the form adopted here,
namely $f_Q =(P/2\pi t_c)^2$, but \citet{rasio96} adopt $f_Q' =
f_Q\pi^2$, i.e.\ larger by almost an order of magnitude.  This factor
translates directly to the implied upper limit on the relative
frequency, weakening the constraint considerably (see Fig.\ \ref{fig:three}).  
Note that this would also weaken the constraint on planets in 
47 Tuc, but by a smaller factor than the constraint on BDs. 

Thus, given the current uncertainties in the physics of tidal
dissipation, I conclude that no robust inferences can made about the
frequency of brown dwarf companions in 47 Tuc, and any conclusions
regarding the universality of the initial mass function are probably
premature.

\section{Implications and Prospects for Detection\label{sec:implications}}

It is clear that, under certain sets of assumptions, extremely close-in
tidally-captured massive planets should exist around stars formed in
dense stellar systems.  What are the prospects for the detection of such
companions, and what would be the implications of any such detections?

Given their extremely close orbits, the most promising method for
detecting companions is transits.  For a planet captured at an orbit
of $a \la 2R_\odot$, the transit probability is ${\cal P}_T \ga 50\%$.
The duration of the transit is
\begin{equation}
t_{T} = \frac{P}{\pi} {\rm arcsin}\left[ \sqrt{{\cal P}_T^2 - \cos^2 i}\right],
\label{eqn:transdur}
\end{equation}
where $i$ is the inclination. Thus for a central transit, the duty
cycle (fraction of time in transit) is quite large: $t_T/P \sim 15\%$.
For a solar-type primary, the transits will last $\sim 50~{\rm min}$
and recur every $\sim 5~{\rm hours}$.  The transit depth will be $\sim
1\%$.  Given the extremely close proximity of the planet to the star,
one might worry that the planet will induce additional photometric
variations on the star.  The reflected light from the companion will
induce an fractional flux variation of order $\Delta F/F \sim
p(R_p/a)^2 \sim 2\times 10^{-3}$, where $p$ is the geometric albedo,
and I have adopted $p=2/3$ as appropriate for Lambert sphere
scattering.  Although five times lower than the signal from the
transit, a signal of this magnitude may be large enough to be
detected.  In fact, depending on the selection criterion used to
choose targets for the transit search in 47 Tuc, this may be cause for
concern in interpreting this result.  The planet will also induce
photometric variations of the star from tidal and Doppler effects,
but these will likely be small for planetary companions
\citep{loeb03}.  They may, however, be significant for more massive
companions.  Thus, barring an unforeseen effects that induce
additional photometric variability, it should be possible to cleanly detect
tidally-captured planets via transits.

Because of their short periods and relatively large duty cycles, the
observational requirements of a transit search for tidally-captured
planets are generally much less severe than transit searches for more
distant planets.  In particular, complete coverage of several phases
can be achieved in less than a few days.  This is especially important
for ground-based transit searches, as the coherence time of weather
patterns is typically several days.  This proves to be devastating
when searching for planets with periods of order the coherence time,
but nearly optimal for very short periods.  Furthermore, aliasing is
minimized because the period of the planet is typically less than the
duration of a typical clear observing night.

Tidally-captured planets will also induce radial velocity variations
in the host star, of amplitude $K\simeq 340~{\rm m~s^{-1}}(5~{\rm
hr}/P)(M_p/\mjup)$.  For reasonably bright sources, this is well
within reach of current instrumentation.  However, the presence of
such a close planet may induce additional radial velocity variability,
making detection of the planetary signal difficult.  Furthermore, the
extremely low expected frequency of tidally-captured planets makes
this an extremely inefficient method of detection, unless multi-object
spectrographs capable of precision radial velocity measurements become
available, so that many stars can be monitored simultaneously.

The systems that offer the best chance of observing tidally-captured
planets are globular clusters, because they have the highest
frequencies of captured planets ($\sim 0.025-0.1\%$), and the largest
total number of systems.  Furthermore, due to the large velocity
dispersions, planets are captured into very close orbits.  This
increases the a priori probability that the planet will transit its
parent stars to nearly unity.  In \S\ref{sec:bds}, I considered in
detail frequency and detectability of tidally-captured planets in 47
Tuc.  Although no transiting planets were found in this cluster, it
may still be worthwhile to search in other clusters.  An observing campaign
with the Hubble Space Telescope need not be as ambitious as that toward
47 Tuc. 

The expected number of transiting, tidally-captured planets in a rich
open cluster is only of order unity.  Therefore, the prospects of
detecting such planets in transit searches toward open clusters
\citep{street03,burke03b} are poor, unless many massive planets are
formed per star.

Loose clusters are subject to fairly rapid disruption, after which
their stars disperse into the general disk population.  It is
possible that a large fraction of stars in the disk were initially
born in such environments.  Assuming a typical lifetime of a loose
cluster of $\sim 300~{\rm Myr}$, then $\sim 0.005\%$ of stars in the
field should have a tidally-captured planet, if all stars in the field
were born in associations, and planet formation were ubiquitous.
Accounting for a binary fraction of 50\%, and assuming a transit
probability of $\sim 40\%$ for planets captured in loose clusters, I
estimate that $\sim 100,000$ stars need to be monitored to have a
chance of detecting even one transiting tidally-captured planet.  OGLE
monitored 5 million stars toward the Galactic bulge over 45 nights to
search for transiting planets to disk stars \citep{udalski02}.  They
only analyze a subset of $\sim 52,000$ of these stars with photometry
better than $1.5\%$.  Similarly, the EXPLORE project has monitored
$\sim 350,000$ stars in the Galactic plane for 11 nights
\citep{mallen03}.  They achieved better than $1\%$ photometry for a
subset of 37,000 stars, which they visually inspected for transits.
In both cases, an insufficient number of stars were searched to place
interesting constraints on the fraction of tidally-captured planets.
However, because of the large number of expected transits, it should
be possible to detect transiting planets even when the scatter is
larger than the expected signal.  Therefore, the optimal (but
time-consuming) approach would be to search all light curves for
transiting planets.  For such a photon-limited survey, the detection
probability scales as $P^{-5/3}$ \citep{pepper03}, highly favoring
extremely close-in tidally-captured planets.

It is important to reiterate that all of these frequency estimates
rely on the assumption that the tidal decay timescale is longer than
$\sim 1~\gyr$, and thus that the quadratic prescription for the
suppression of eddy viscosity is approximately correct.  The only
exception is planets captured in loose associations.  Such planets are
captured sufficiently far from their parent stars that they are stable
over $\sim~\gyr$ timescales for both linear and quadratic
prescriptions.  However, for ages much larger than $1~\gyr$, the orbits
will rapidly decay under the linear prescription (see Fig.\
\ref{fig:two}).  The detection of a planet on an extremely close orbit
would therefore have important implications for the tidal dissipation
theory.

How can tidally-captured planets be distinguished from planets that
migrated to such close orbits?  The period distribution of
radial-velocity planets shows a pile-up at $P\sim 3~\days$, with
a significant lack of planets with smaller periods.  Tidally-captured
planets would be easily identified in such a distribution.
\citet{kandl02} suggest that this distribution arises from the fact
that the centers of protoplanetary disks are evacuated interior to
separations corresponding to periods of $6~\days$, and that
planetary migration thus halts when the planet's outer Lindlbad
resonance reaches the inner disk edge.  Although this seems to provide
a compelling explanation for the observed period distribution of
radial velocity planets, the recent detection of a transiting planet at
$P=1.2~{\rm days}$ \citep{konacki03} generally suggests that this
picture cannot be universal \citep{sasselov03}.  Indeed,
\citet{trilling98} suggest another mechanism for halting migration:
planets migrate until they reach their Roche limit, at which point
they lose enough mass to the parent star to balance inward migration.
If disk dispersal occurs in a sufficiently short time, the planets
will be left intact in short-period orbits.  Such planets would
generally have similar periods as tidally-captured planets.
Therefore, additional diagnostics are needed.

One possible method of distinguishing tidally-captured planets from
migrating planets is a radius measurement.  Both known transiting
planets have radii significantly larger than that of Jupiter.  It has
been suggested that these large radii are due to the fact that these
planets migrated on a sufficiently short timescale that their
gravitational contraction was retarded, leaving them in a permanently
inflated state \citep{burrows00}.  If this process is universal, then
we can expect migrated planets to have significantly larger radii than
tidally-captured planets, which could not have migrated far from their
birthplaces.  This assumes that there are no processes that can
inflate planets once they have been tidally captured.  In principle,
variable tidal forces from the star on a planet in an eccentric orbit
can deposit enough energy to inflate the planet \citep{gu03}, but to
have a significant effect, there must be some external force
continuously pumping the eccentricity, since the circularization
timescale for extremely close-in planets is very short.

\section{Summary}

Planets which form around stars born in dense stellar environments are
subject to dynamical perturbations from other stars in the system.
These perturbations will generally serve to strip outer planets from
their parent stars, leading to a significant population of
freely-floating planets in the system.  Further interactions with
stars in the system drive this planetary population toward
equipartition, thus raising the velocity dispersion of the planets,
ultimately resulting in escape from the system.  However, some
planets will survive long enough to be pass sufficiently
close to another star to be tidally captured.

The stability of tidally-captured planets against orbital decay
depends critically on the physics of viscous dissipation of the tide
in the convective envelope of the star.  This processes is extremely
uncertain, and different prescriptions lead to decay timescales that
differ by four orders of magnitude.  For quadratic suppression of eddy
viscosity, planets on tidally-captured orbits will generally be stable
for $\ga \gyr$, whereas for linear suppression, planets will decay
very quickly.

The frequency of tidally-captured planets depends on the competing
effects of dynamical stripping, evaporation, capture, and orbital
decay.  The timescales for these effects depend in turn on the
properties of the stellar systems, particularly the velocity
dispersion $\sigma$ and the number density $\nu$.  Under the
assumption that all stars form four Jupiter-mass planets 
with a uniform logarithmic distribution in semi-major axis between $5-50\au$, and
assuming quadratic suppression of eddy viscosity, I have estimated the
frequency of tidally-captured planets in a stellar system as a
function of the age, velocity dispersion, and number density of the
system.

For loose associations with $\sigma \sim 0.6~\kms$ and $\nu\sim 10^2~\pcpc$, 
the frequency of tidally-captured planets after a typical cluster
lifetime of $10^8$ years is $\sim 0.005\%$.  If most stars formed in
such systems, this is roughly the expected frequency of extremely
close-in planetary companions to stars in the field.  These planets may be found by
deep, wide-angle searches for transiting planets around disk stars
\citep{udalski02, mallen03}.  Detection of such a population would
imply both that most stars were formed in dense stellar environments,
and that the timescale for orbital decay is $\ga \gyr$.

For rich open clusters ($\sigma \sim 1.5~\kms$ and $\nu\sim 10^3~\pcpc$), 
the frequency of tidally-captured planets is $\sim 0.03\%$.
Unfortunately, the prospects for detecting such a population of
planets in a given cluster are poor, as $\la 1$ transiting planet
per cluster is expected.

Due to the long relaxation times and high encounter rates, the
frequency of tidally-captured planets in a typical globular cluster
($\sigma \sim 10~\kms$ and $\nu\sim 10^4~\pcpc$) is $\sim 0.1\%$.  Therefore,
one expects several hundred transiting planets in each globular
cluster.   

The relatively high frequency of tidal capture in globular clusters
makes them excellent targets to search for such systems.  The null
result of the search for transiting planets in 47 Tuc can be used
to place interesting constraints on the initial frequency of planets
and brown-dwarfs (either initially free-floating or bound)
relative to stars.  I find that $\la 1$ planet with mass $\ge \mjup$
originally formed per star in 47 Tuc.  Less than 50\% of stars formed
solar-system analogs.  The initial frequency of brown dwarfs relative
to stars is $25\%$.  All of these conclusions are under the assumption
of quadratic suppression of eddy viscosity.  If viscosity suppression
is much less efficient, no interesting constraints can be placed.

Tidally-captured planets will have orbital periods of $\sim
0.1-0.4~\days$, and can be distinguished from migrating planets
from their distribution of orbital periods provided the mechanism that
is observed to halt planetary migration at $P\sim 3~\days$ is
robust. However, if the recent discovery of a planet at $P\sim 1~\days$ 
is indeed indicating that there exist multiple mechanisms for
halting migration, then one cannot rule out the hypothesis that an
observed extremely close-in planet was the result of orbital
migration.  However, the radii of tidally-captured planets are
expected to be similar to their radius at their sites of formation,
and thus considerably smaller than planets that migrated quickly to
close orbits.  Thus a radius measurement might provide a diagnostic
for tidally-captured planets.

If the majority of stars is formed in dense stellar environments,
then most planetary systems will be subject to a variety of dynamical
effects that can alter planetary formation, migration, and survival.
Therefore, planetary systems cannot be understood as isolated systems, and the
interpretation of the observed properties of extrasolar planets will
require the consideration of these effects.  This study highlights
one example of how such considerations can lead to
new observable diagnostics of planetary formation.

\acknowledgments I would to thank John Bahcall, Ian Bonnell, Neal Dalal, and Roman
Rafikov for comments and discussions. This work was supported by NASA
through a Hubble Fellowship grant from the Space Telescope Science
Institute, which is operated by the Association of Universities for
Research in Astronomy, Inc., under NASA contract NAS5-26555.

\end{document}